\documentclass[a4paper,11pt]{article}
\usepackage[a4paper,includemp=false,body={16cm,24.7cm},vcentering]{geometry}
\usepackage[latin1]{inputenc}
\usepackage[T1]{fontenc}
\usepackage{mathptmx} % police Times texte et maths
\usepackage[bf,center]{caption}
\usepackage{graphicx} % si figures

\usepackage{hyperref,amsmath,amssymb,amsthm}

\usepackage{multirow}

\makeatletter
\renewcommand{\section}{\@startsection{section}{1}{0mm}{30pt}{12pt}{\normalfont\normalsize\bfseries}}

\renewcommand{\subsection}{\@startsection{subsection}{2}{0mm}{18pt}{12pt}{\normalfont\normalsize\itshape}}
\makeatother
\newcommand{\Title}[1]{\begin{center}{\bfseries\fontsize{12pt}{12pt}\selectfont#1}\end{center}}
\newcommand{\Author}[2]{\begin{center}{\fontsize{12pt}{12pt}\selectfont#1}\\{\it #2~}\end{center}}

\newcommand{\Conclusion}{\section*{Conclusion}}

\begin{document}

\Title{\Large{Photometric Science Alerts from Gaia}}
  
\Author{{\L}ukasz Wyrzykowski$^{1,2}$, Simon Hodgkin$^2$, Nadejda Blogorodnova$^2$, Sergey Koposov$^2$, Ross Burgon$^3$}{1. Astronomical Observatory of the University of Warsaw, Al. Ujazdowskie 4, 00-478 Warszawa, Poland, email: lw@astrouw.edu.pl\footnote{name pronunciation in English: {\it woocash vizhikovski}}\\
2. Institute of Astronomy, University of Cambridge, Madingley Road, CB3 0HA, Cambridge, United Kingdom, \\
3. Department of Physical Sciences, The Open University, Walton Hall, Milton Keynes, MK7 6AA, United Kingdom}

\subsubsection*{Abstract}
Gaia is the cornerstone mission of the European Space Agency. From late 2013 it will start collecting superb astrometric, photometric and spectroscopic data for around a billion of stars of our Galaxy. While surveying the whole sky down to V=20mag Gaia will be detecting transients and anomalous behaviour of objects, providing near-real-time alerts to the entire astronomical community. Gaia should detected about 6000 supernovae, 1000 microlensing events and many other interesting types of transients. Thanks to its on-board low-dispersion spectrograph the classification of transients will be robust, assuring low false-alert rate. 
We describe the operation of the Photometric Science Alerts system, outline the scientific possibilities and conclude with an invitation to collaborate in the ground-based follow-up Gaia alerts during the early months of the mission when the outcome of the alerting pipeline needs to be verified. 
\newline
\newline
{\it Based on an invited talk presented at the Gaia-FUN-SSO-2 International Workshop, Paris Observatory, 19-21 September 2012.}

\section{Introduction}

\noindent

ESA's Gaia mission will be launched in late 2013 and will observe the entire sky for 5 years providing ultra-precise astrometric measurements (positions, parallaxes and proper motions) of a billion stars in the Galaxy.  The astrometry will be derived from multiple observations of each source at different scanning angles.  Hence, naturally, Gaia becomes an all-sky multi-epoch photometric survey, which will monitor and detect variability with millimag precision down to $V=15$ mag and about 0.01 mag precision down to $V=20$mag.  Gaia will also be able to detect new objects appearing in its field-of-view thanks to the window allocation system in the first CCDs in the focal plane.  This includes most classes of transient phenomena like supernovae, novae, microlensing events, asteroids, etc.

In the first part of this proceedings we describe the design of the Gaia Science Alerts system, which will run daily at the Institute of Astronomy in Cambridge, UK, and is responsible for the detection and classification of photometric transients.  In the second part we briefly describe potential scientific opportunities related to Gaia alerts.  Thirdly, we present the status of preparations for the alert verification phase in the early days of the mission.  We conclude with an invitation to collaborate in the ground-based follow-up Gaia alerts.

\section{Gaia Science Alerts system}

\subsection*{Data flow}
\noindent
The alerting system, called AlertPipe, will be run on a daily basis at the Institute of Astronomy in Cambridge (part of the Gaia Data Processing and Analysis Consortium, DPAC). Gaia satellite will reside in the 2$^{\rm nd}$ Lagrange Point (L2) and the data gathered during the scanning of the sky will be downlinked to the ground every day during an 8h window of visibility. The data will be then transferred to DPAC nodes in Germany and Spain where it will be pre-processed during the Initial Data Treatment (IDT) process.  One of the main tasks of the IDT is to crossmatch all star detections with previous Gaia detections (in early days of the mission it will rely on the Initial Gaia Source List, a catalogue compiled from a variety of existing ground-based observations).  As soon as IDT finishes processing a data packet, containing typically about 50 million objects, it will be transmitted to Cambridge and analysed by the AlertPipe.  The total lag between an observation and the AlertPipe processing and alerting is expected to vary from a few hours to 48 hours, and depends on many factors, including the region of the sky (high stellar density regions lead to a delay in downlink for faint targets) and brightness of the object (downlinking order depends on the brightness, however not linearly). 
AlertPipe processing is expected to take no more than couple of hours, depending on the amount of data to process and size of the historic database. 

\subsection*{Gaia sampling}
\begin{figure}
\centering
\includegraphics[width=0.52\textwidth]{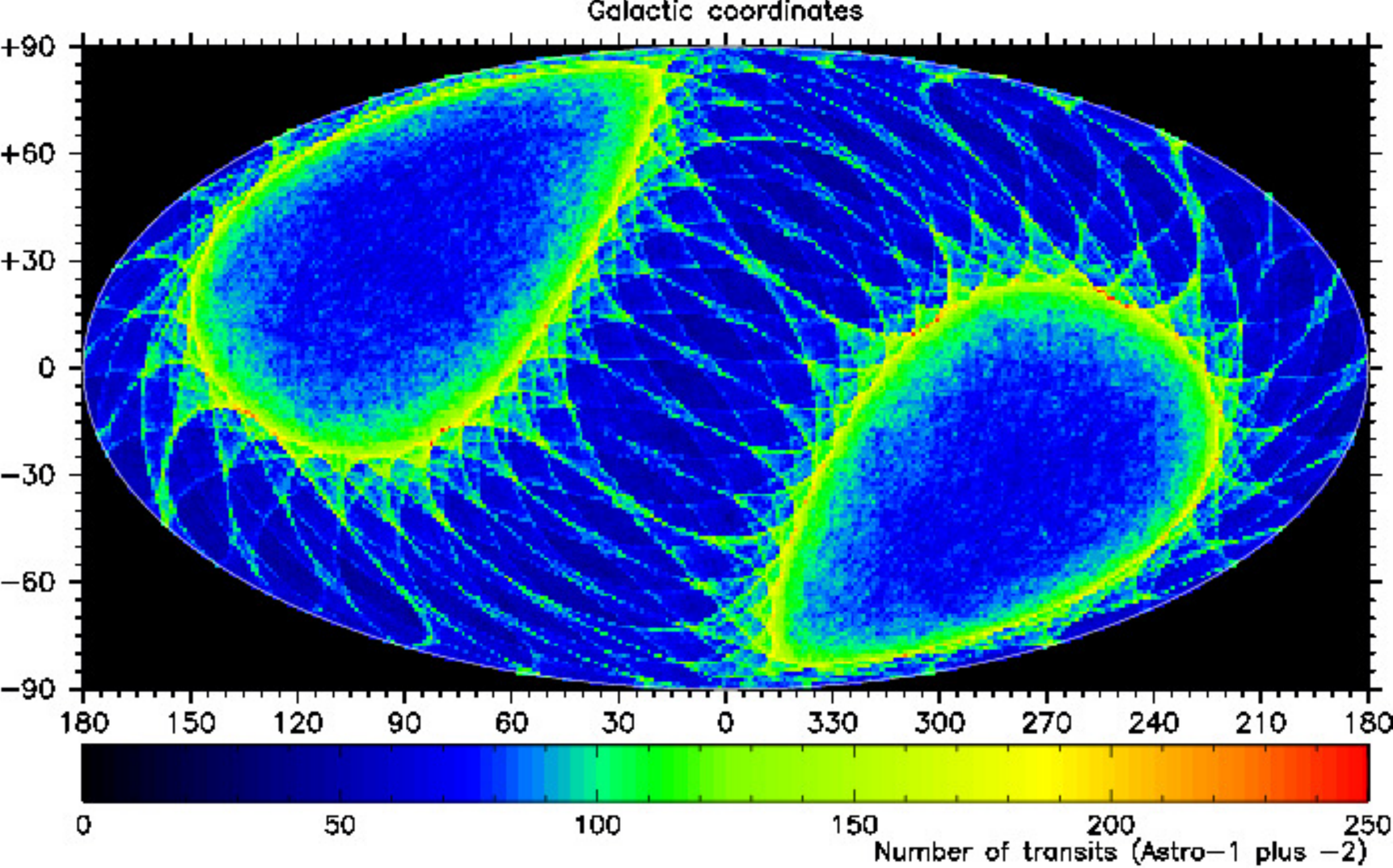}
\includegraphics[width=0.47\textwidth]{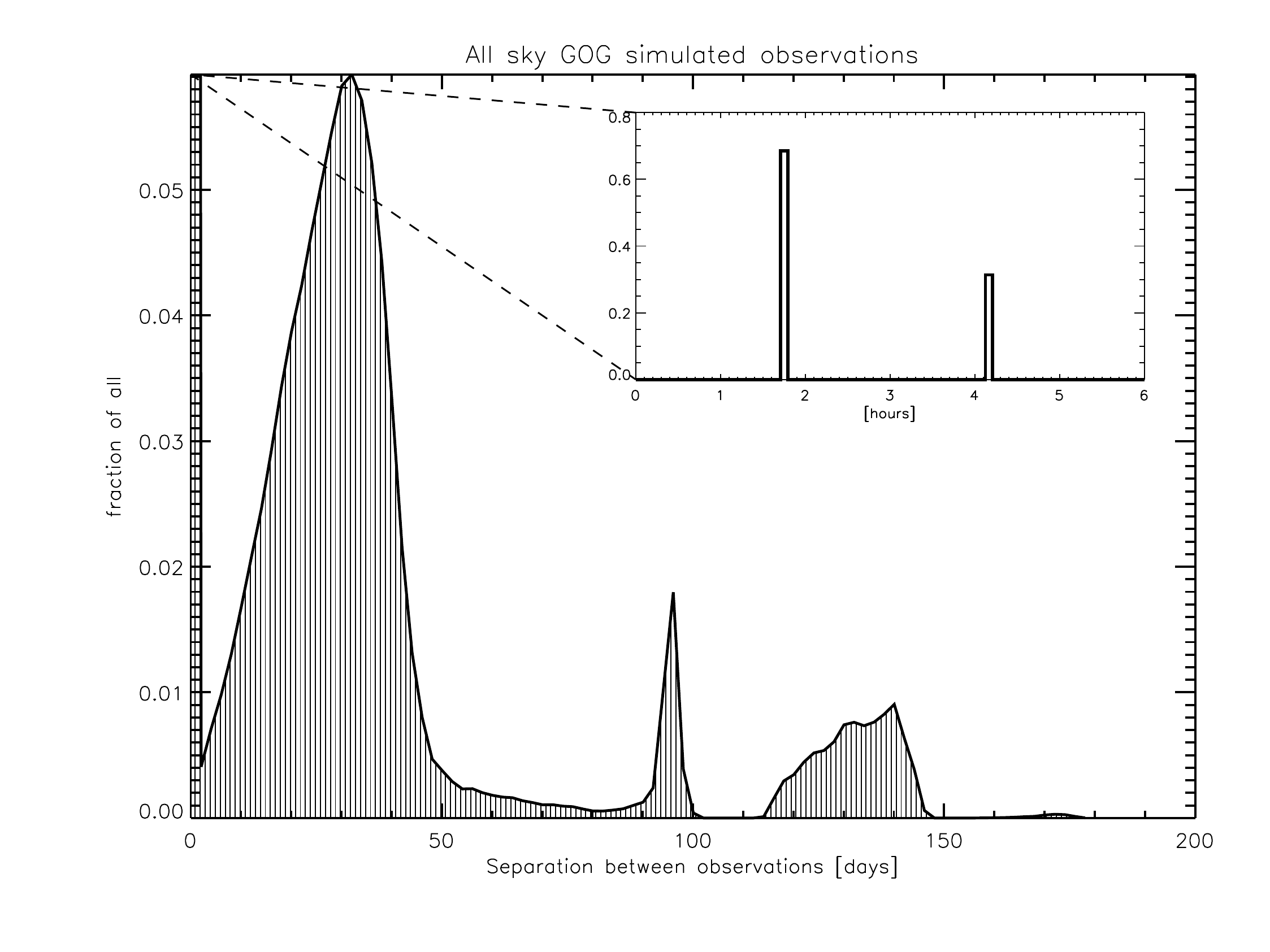}
\caption{Left: Sky scanning pattern of Gaia showing the number of observations for each object in Galactic coordinates. Regions of the Galactic Centre will have about 50 data points, whereas high density regions, at ecliptic latitude 45 and -45, will have about 200 measurements. Right: Time sampling pattern of Gaia (time difference between two subsequent observations) is dominated by the 106 mins separating the two telescopes. The Gaia spacecraft spins once every 6 hours.} 
\label{fig:scanning}
\end{figure}

\noindent
The observing strategy across the sky, called the Nominal Scanning Law (NSL), is a pre-defined pattern, optimised for the final astrometric solution (Lindegren et al. 2012) \cite{Lindegren2012}, and ensures that most of the stars will obtain, on average, about 70 measurements at different scanning angles.  Some areas of high stellar density such as the Galactic Bulge will only have about 50 measurements during the entire mission. On the other hand, regions within a few degrees of ecliptic latitudes of $\pm$45 degrees will be scanned approximately 200 times, see Fig.  \ref{fig:scanning}

Gaia consists of two 1.45x0.5m mirrors set at angle of 106.5 degrees perpendicular to a slowly precessing spin axis. One full rotation of the satellite takes exactly 6h, therefore the preceding and following fields-of-view will observe the same patch of the sky with a separation of 106.5 minutes. After this pair of observations there may come another pair (and many other pairs at the ecliptic nodes at b$\pm$45 deg), but due to the precession of the spin-axis, in most cases the fields-of-view will quickly precess out of that area of the sky. Typically, the same field will be observed again after 30 days or more, see Fig. \ref{fig:scanning}.

\subsection*{Detection}
\noindent
The anomaly detection system within the AlertPipe depends on the crossmatch information from the IDT such that sources not matched with known objects are flagged as ``new''. All new objects passing a detection threshold (early in the mission set to about Gaia broad filter magnitude G=19) will be first checked against possible asteroid positional coincidence\footnote{The asteroids are being recognised from these new IDT sources by a parallel alerting system running within DPAC, see papers by W.Thuillot and P.Tanga from this workshop.} and all surviving candidates for new transients will pass to the next stage, the classification.

Known sources, namely those with some prior observations, will be checked to see if the new observations are in any way anomalous with respect to the data gathered so far by Gaia. The AlertPipe stores all observations of all objects and at this stage runs various detection algorithms in order to check against anomalies, for example, mean--{\it RMS} detector or simple {\it delta-magnitude} threshold detector. Each of the detectors is sensitive to a different kind of anomalies and some are more suitable for different stages of the mission, depending on the amount of historical data available. The thresholds of the detectors are tuneable during the entire mission and will evolve with the increasing amount of gathered data and better understanding of the instruments.

The focal plane of Gaia contains 7 rows with 9 CCD each\footnote{Central row contains only 8 photometrically useful CCDs, as the last one is replaced by the wavefront sensor CCD.}, on which the brightness and position of an object is measured during a scan. This means that a single transit will contain 9 data points, each separated by about 4.4 seconds. This allows not only for immediate ruling out of cosmic rays and other instrumental artefacts affecting the photometry, but also for testing the short-term variability of any source.

\subsection*{Classification}
\noindent
The next stage of the AlertPipe data processing is responsible for the preliminary classification and filtering of detected candidates for alerts. This will employ both photometry and low-resolution spectroscopy from Gaia. The Blue and Red prism-based Photometers (BP/RP), installed on the focal plane after the astrometric CCDs, will cover spectral ranges 330-680nm and 640-1000nm, respectively, with resolution of R$\sim$100. 

Most of the Gaia observations will come in pairs separated by 106.5 minutes and in case of most transients the downlinked daily portion of data will contain both observations. This will not only provide a double check on the possible transient candidate, but is also suggestive of a light curve classifier which can exploit the flux-gradient as an indicator of object type. For example, a simple slope-amplitude Bayesian classifier can provide a probability distribution for a transient being a cataclysmic variable, supernova or long period variable, popping up from the background. Tests of such classifier performed on the SDSS Stripe82 and OGLE data have shown that with just two data points we are able to distinguish between these major types of transients with relatively high accuracy.

%\begin{figure}
%\centering
%\includegraphics[width=0.55\textwidth]{classifier-slope-ampl.pdf}
%\caption{Slope-amplitude diagram with training set for the light curve classifier. The Bayesian classifier returns a probability distribution for an event described by just two data points. } 
%\label{fig:classifier}
%\end{figure}

Many more `features' are available to aid with classification, including the BP/RP spectra. Simulations with Gaia BP/RP spectra for Supernovae have shown that most detections by Gaia can be further subclassified by type, epoch and redshift for transients brighter than 19 mag. This unique capability of Gaia will help to improve our classifications, and will allow for more targeted high-resolution spectroscopic follow-up.
 
\begin{figure}
\centering
\includegraphics[width=0.43\textwidth]{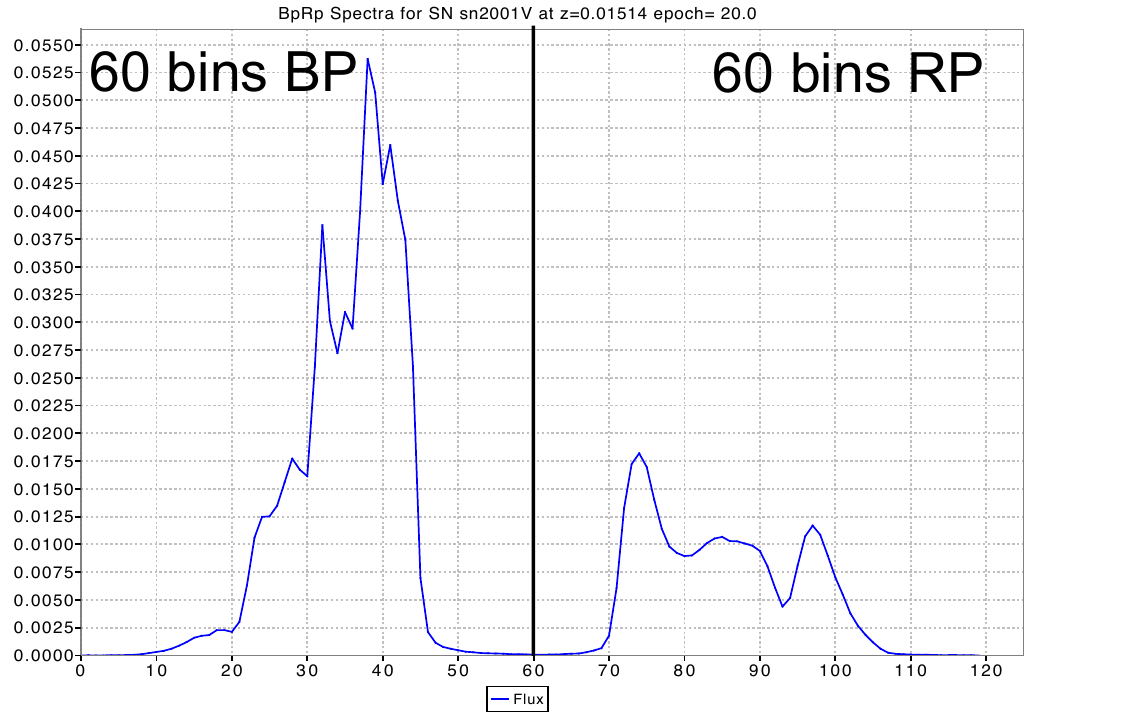}
\includegraphics[width=0.56\textwidth]{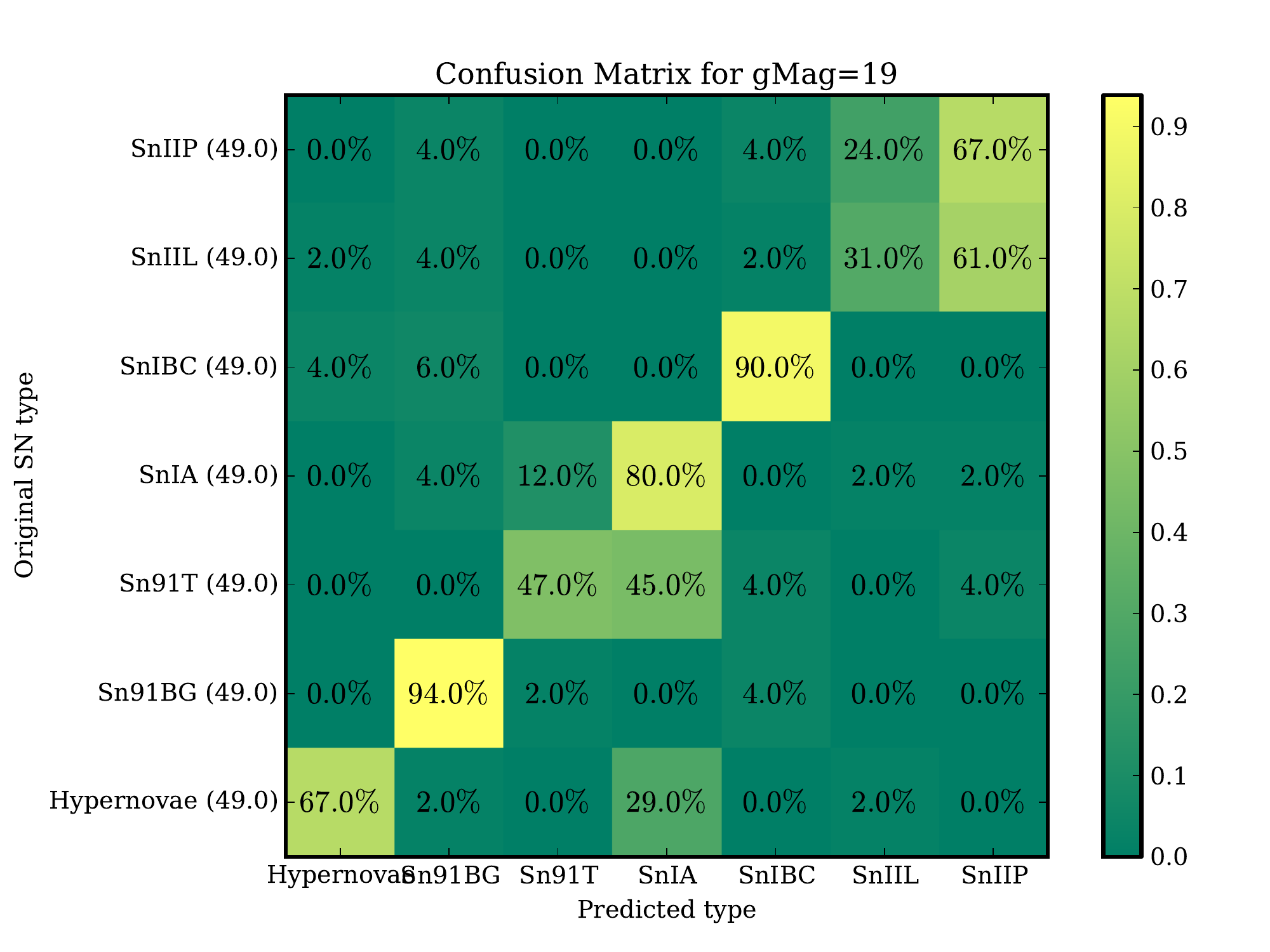}
\caption{Left: the spectrum of type Ia supernova as seen by Gaia's Blue and Red Photometers (BP/RP). The resolution of about R$\sim$100 is enough to unambiguously derive the type, epoch and redshift of most supernovae down to $\sim$19 mag. Right: confusion matrix for supernovae BP/RP spectra classification at 19 mag. Most major types are recognised with relatively low confusion. } 
\label{fig:SNspectrum}
\end{figure}

The filtering and classification of the transient events will be supplemented by contextual information obtained from available archival catalogues, for example, 2MASS, SDSS, ASAS, OGLE, HST, VISTA and so on. Known variable stars will be crossmatched with candidate transients and supernova or tidal disruption event classification will be enhanced if a galaxy can be associated with the event. We will also cross-match our candidate alerts with recent alerts reported by other transient surveys, on-going during the Gaia mission. 

\subsection*{Data dissemination}
All Gaia alerts will be public immediately after discovery and preliminary classification. The alerts will be disseminated to the astronomical community via a number of protocols, including traditional email and web server, as well as machine readable means like VOEvents.  Each alert will provide coordinates of an event, a light curve collected so far by Gaia, BP/RP spectrum and the results of the cross-match and classification analysis.

Early in the mission, during the verification phase, the alerts from Gaia will only be available to a dedicated team of telescopes and astronomers involved in the alerts verification, to assure the robustness of the detection and classification pipeline (see more below).

\section{Scientific opportunities}
The scanning law and the time lag between observation and analysis makes the Gaia transient survey more sensitive to longer events (those lasting from a couple of hours to months). These include: supernovae, dwarf novae, classical novae, microlensing events, tidal disruption events, R CrB-type stars, FU Ori-type stars and Be stars.

\subsection*{Supernovae and other extragalactic transients}
In its unbiased search for supernovae, Gaia will be capable of detecting in total about 6000 SNe brighter than G=19 (10000 to G=20) (Altavilla et al. 2012 \cite{Altavilla2012}, Belokurov\&Evans, 2003 \cite{BelokurovSNe}). One third of those will be detected before maximum, which will allow for detailed follow-up and studies of supernova evolution. With 3--4 supernovae discovered every day, it will require well-organised follow-up, as the Gaia data alone will not be enough to provide sufficiently detailed light curves, e.g. needed for cosmological applications of the SNe.

As mentioned above, Gaia will provide an auto-follow-up of its own targets with low-resolution spectra available for every source.  This will also allow for rapid recognition of unusual and rare types of supernovae, for example, Super Luminous SNe (Quimby et al. 2011) \cite{Quimby2011}, which reach -23$^{\rm{rd}}$ mag (absolute), or Luminous Red Novae, which bridge the gap between classical novae and supernovae (Kasliwal et al. 2011) \cite{Kasliwal2011}.

Gaia will not be best-suited for real-time discoveries of optical afterglows of Gamma Ray Bursts. Nevertheless, simulations predict that it should be able to detect about 20 brighter and longer GRBs and Orphan Afterglows (Japelj \&Gomboc 2011) \cite{Japelj2011}, for which timely follow-up would be critical.

\begin{figure}
\centering
\includegraphics[width=0.6\textwidth]{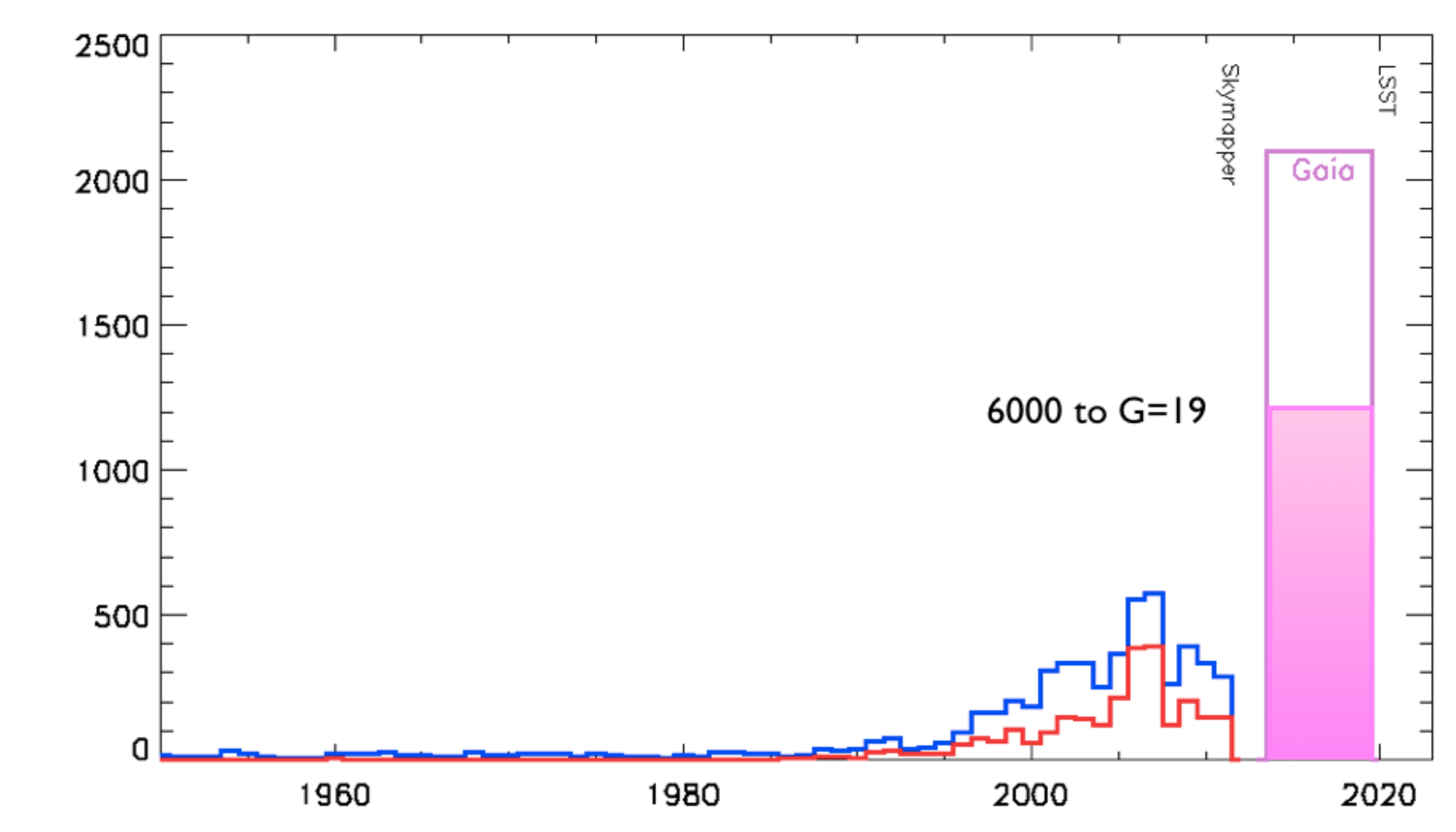}
\caption{Supernovae detection rate per year. Gaia is expected to detect 6000 SNe down to 19 mag, with 1/3 of them discovered before the maximum brightness. With the detection threshold at 20 mag Gaia might see even 2100 supernovae a year. } 
\label{fig:supernovarates}
\end{figure}

\subsection*{Microlensing Events}
Gravitational microlensing events, the temporal brightening of distant stars due to the mass of a foreground lens passing in front, typically occur in the densest regions of the sky, i.e. the Galactic Bulge and the Galactic Plane. About 2000 such events are currently being detected every year by the dedicated microlensing surveys OGLE (Udalski 2003) \cite{UdalskiOGLE} and MOA (Sumi et al. 2003)\cite{SumiMOA}. Among the lensing systems more than a dozen planets have been found, e.g. Gaudi et al. (2008) \cite{Gaudi2008}. Significant results would also include detections of brown dwarfs and black holes, studies of stellar atmospheres and investigations of the structure of the Galaxy.

The central regions of the Galaxy will continue to be monitored by the OGLE and MOA surveys during the Gaia mission, however, any microlensing events found outside of these fields will require an intense photometric follow-up in order to make them scientifically useful. We expect Gaia to detect about 1000 microlensing events from all over the sky. 

Interestingly, Gaia's astrometry will be able to provide very precise measurements of the lensed source displacements, caused by microlensing. Combination of Gaia's astrometry and photometry, with more densely sampled ground-based photometry could lead to the derivation of the mass of the lens, including discoveries of lenses from the stellar remnant population, like neutron stars or black holes.

\subsection*{R CrB-type stars}
These mysterious stars exhibit spontaneous and unpredicted dimmings in their light curves by as much as 8 magnitudes.  Catching such events ``red-handed'' and triggering detailed photometric and spectroscopic studies would allow for a much better understanding of the nature of these stars, thought to be linked with the stellar mergers. This is turn may help decipher supernova progenitor models.  There are about 50 R CrB-type stars known in our Galaxy, found mainly during large scale photometric campaigns of MACHO, OGLE and ASAS surveys, e.g. Tisserand et al. (2011) \cite{TisserandRCB}.  Gaia should find many new examples of these stars at fainter magnitudes from all over the sky. BP/RP spectroscopy will help with the classification of these objects.

\subsection*{Other types of transients and the Watch List}
Gaia will detect numerous cataclysmic variables, located primarily in the Galactic Plane, providing a uniform large sample of these objects.
We expect also to alert on very rare events like outbursts FU Ori- or EX Lup-type young stars, of which only a handful is known. 
Outbursts of Be-type stars will also be detected by Gaia alerting system. There should be up to 600 events during the entire mission for Be stars brighter than 12 mag, giving an opportunity for detailed high-resolution spectroscopy of these objects during outbursts. 

We plan also to provide near-real-time Gaia photometry of a limited number of selected interesting targets, known for their erratic behaviour. 
For example, a known FU Ori-type star can be on the Watch List and whenever it is observed by Gaia, the most up-to-date brightness measurement will be made available, allowing for continuous monitoring of the object.

\section{Alerts Verification}
Gaia is expected to be launched from French Guyana in the end of 2013. It will take about two months to reach the L2 and another two months to get all the systems fully operational. For the next 1 or 2 months it is planned that Gaia will operate in a special Ecliptic Poles Scanning mode, during which both Ecliptic Poles will be intensively observed (the satellite precession period will be slowed down significantly). During this stage the data analysis systems will be vigorously tested, including the alerting system.  After that period, Gaia will start its regular scanning of the entire sky, but the alerting system will be turned on gradually, waiting for enough observations to be gathered for a reference for detections of transients. As soon as there is enough data collected for some regions of the sky, for example, requiring at least 10 prior observations, the alerting system will start operating.

It is ESA's policy that all Gaia data become publicly available. However, before the alerts start flowing to the astronomical community, the detection and classification system has to be verified to assure good quality and robustness of public alerts.  Therefore, for the first couple of months of the functioning of the alerting system we envisage a Gaia Alerts Verification Phase, during which the end-to-end alert system will be thoroughly shaken down by a dedicated team of astronomers working in collaboration with the AlertPipe developers. Some aspects of the verification can start during the Ecliptic Poles scanning, for example classification of large numbers of known variable stars (Soszy{\'n}ski et al. 2012) \cite{SoszynskiSEP}. The Verification Team (VT) will see some of the first Gaia data and will have a chance to prepare for the follow-up of the Gaia alerts when they become openly available.

Therefore we issue an invitation to get involved in the early verification of the Gaia Alerts system. Verification observations will validate the outcome of the alerting pipeline and will involve e.g.: confirming the alert, building a detailed multi-band light curve and obtaining a spectrum. Because the magnitude range of Gaia alerts is wide (V=5--20) we welcome observers equipped with telescopes of any size. Longitudinal and latitudinal coverage on the globe will be needed in order to assure accessibility and long-term visibility of the targets.

Potential members of the VT, should pass a couple of infrastructure tests, including performing a number of follow-up observations of alerts from current surveys, e.g. Catalina Real-Time Transient Survey\footnote{http://crts.caltech.edu}. The data should be reduced promptly and submitted for verification to the central repository. The verification network is currently being formed and should be ready for operation by the end of 2013. For more details visit the pages of Gaia Science Alerts Working Group\footnote{http://www.ast.cam.ac.uk/ioa/research/gsawg/} or contact the authors.

\Conclusion From mid-2014 the Gaia mission will begin to deliver near-real-time alerts on anomalous or transient events from the entire sky down to V=20. The G-band photometry and BP/RP low-dispersion spectrometer will allow early detailed classification, and in the case of supernovae will also provide an estimate of the redshift and epoch. Before the alert stream becomes publicly available, a period of verification will take place in the first months of the mission. For this we encourage an involvement from small and large telescopes from around the globe.

\section*{Acknowledgements}
The hard work and help received from numerous members of the Gaia DPAC is acknowledged here. {\L}W acknowledges support from "Iuventus Plus" programme of the Polish Ministry of Science and Higher Education, award no IP2011 062371.

\end{document}